\documentclass[12pt]{spieman}  
\usepackage{amsmath,amsfonts,amssymb}
\usepackage{graphicx}
\usepackage{setspace}
\usepackage{tocloft}

\title{Estimation and correction of the instrumental perturbations of VBT Echelle spectrograph using a  model-based approach}

\author[a,*]{Sireesha Chamarthi}
\author[a]{Ravinder K Banyal}
\author[a]{S. Sriram}
\affil[a]{Indian Institute of Astrophysics, Bangalore, India}

\cftpagenumbersoff{figure}
\cftpagenumbersoff{table} 
\begin{document} 
\maketitle

\begin{abstract}

The Echelle spectrograph operating at Vainu Bappu Telescope (VBT), India, is a general purpose instrument used for many high-resolution spectroscopic observations. A concerted effort is being made to expand the scientific capability of the instrument in emerging areas of observational astronomy. The present study is aimed at evaluating the feasibility of the spectrograph to carry out precision Radial Velocity (RV) measurements. In the current design, major factors limiting the RV precision of the spectrograph arise from the movable grating and slit; optical aberrations; positional uncertainty associated with optomechanical mounts and environmental and thermal instabilities in the spectrograph room. RV instabilities due to temperature and pressure variations in the environment are estimated to vary around  120 $ \textrm{ms}^{-1} $ and 400 $ \textrm{ms}^{-1} $ respectively. The positional uncertainty of the grating in the spectrograph could induce a spectral shift $\sim1.4\,\textrm{kms}^{-1} $ across the Echelle orders. A Zemax model is used to overcome the uncertainty in the zero-positioning and lack of repeatability of the moving components. We propose to obtain the ThAr lamp observations and using the Zemax model as the reference, predict the drifts in the positions of the optical components. The perturbations of the optical components from the nominal position are corrected at the beginning of the observational run. After a good match is obtained between the model and the observations, we propose to use a Zemax model to improve the wavelength calibration solution.  We could match the observations and model within $\pm$~1 pixels accuracy after the model parameters were perturbed in a real-time setup of the spectrograph. In this paper, we present the estimation of the perturbations of optical components and the effect on the RV obtained.

\end{abstract}


{\noindent \footnotesize\textbf{*}Sireesha Chamarthi,  \linkable{sireesha@iiap.res.in} }

\begin{spacing}{2}   

\section{Introduction}
\label{sect:intro}  

To utilize a high-resolution spectrograph for precision Doppler spectroscopy, a robust instrument design is essential for the desired performance. A dedicated RV spectrograph has several unique features such as fixed optomechanical components, optics fabricated with low expansion material. Along with this a vibration-free vacuum enclosure is essential to enforce a tight control of environmental conditions (temperature and pressure, etc) \cite{HARPS, AFOE, spectrographs}. Furthermore, the star-light from the telescope is fed through optical fibers for image scrambling and illumination stabilization. Along with the advancement in enhancing the stability and repeatability of the spectrograph, there is significant progress in the wavelength calibration of the instrument. The RV instruments are often equipped with a variety of calibration sources producing well-known reference lines that can be traced to fundamental laboratory standards. HARPS is one such spectrograph that has achieved a wavelength calibration repeatability of 15 $\textrm{cms}^{-1} $ \cite{HARPS_repeatability}.  In recent years, there is also an advancement in the calibration procedure from the empirical method to physical modeling of the instrument \cite{model_based}. Several instrument teams have developed improved calibration procedures by incorporating the physical model of the instrument in the wavelength calibration \cite{FOS, STIS, CRIRES, HESP}. 

To achieve the desired RV precision in a spectrograph,  many of the features mentioned above are incorporated at the design stage itself. Precision in RV measurements is required in many science cases like exoplanet detection, characterization of binary and pulsating stars, analyzing the variability of fundamental constants, exoplanet transit follow- up studies, astro-seismology, etc. Some of these observations require a considerable amount of the telescope time and requires adaptable response time from the telescopes for follow-up observations. In this context, there is a growing demand to utilize the existing spectrographs for these follow-up studies \cite{two_meter,existing_spectrograph}. This can save the telescope time on the large telescopes and adapt the existing instruments for precision RV measurements. The first step in utilizing the existing spectrographs is to critically evaluate the technical and operational level limitations of the instrument and develop a mechanism/model to address them.

Here we have estimated the limitations of a general purpose echelle spectrograph operating on the 2.3 $\textrm{m}$ Vainu Bappu Telescope (VBT) in Kavalur, India for precision RV capabilities. The telescope is equipped with a fibre-fed high resolution (R = 60,000), Echelle spectrograph \cite{VBTechelle}. Since its commissioning in 2003, the instrument is used in a variety of spectroscopic observations of solar system objects, chemical evolution, age and abundance determination of stars and galaxies. However, the spectrograph was not designed originally for the precision RV measurements \cite{stabilityVBT}. For such a general purpose spectrograph, it is essential that the zero-point variations must be accurately tracked and corrected.  In this paper, we have investigated the instability of the spectrograph subjected to perturbations of optical components arising from its current functionality. From the analysis, we propose a methodology for the correction of the limitations from the spectrograph for RV measurements.

 A Zemax model is used for prediction and correction of the positional variations of the components. The effect of the perturbation of optical components on the obtained RV was earlier demonstrated in the works of Gibson \& Wishnow (2016).\cite{synthetic_spectrum}. In the view of increasing the computational flexibility, we analyzed the RV sensitivities of the spectrograph and modeling of observations in Zemax using Python for input and output operations. Invoking the capabilities of Zemax and utilizing it for the spectrograph sensitivity analysis using other software has been demonstrated in previous studies \cite{zemaxmatlab, HERMES}. The known limitations due to the design and usage of the spectrograph for precision RV studies are discussed in Section \ref{sect:limitations}. The approach we have taken address these limitations are presented in Section  \ref{sect:need}. The sensitivity analysis, procedure, and estimation of RV uncertainties are presented in Section \ref{sect:sensitivity}. Matching of the Zemax model with observations is described in Section \ref{sect:model}. Inferences and conclusions of this study are discussed in  Section \ref{sect:conclusion}.

\section{Spectrograph design limitations for precision RV studies}
\label{sect:limitations}  

\subsection{Spectrograph Description}

\begin{figure}
   \begin{center}
   \begin{tabular}{c} 
   \includegraphics[scale = 0.55]{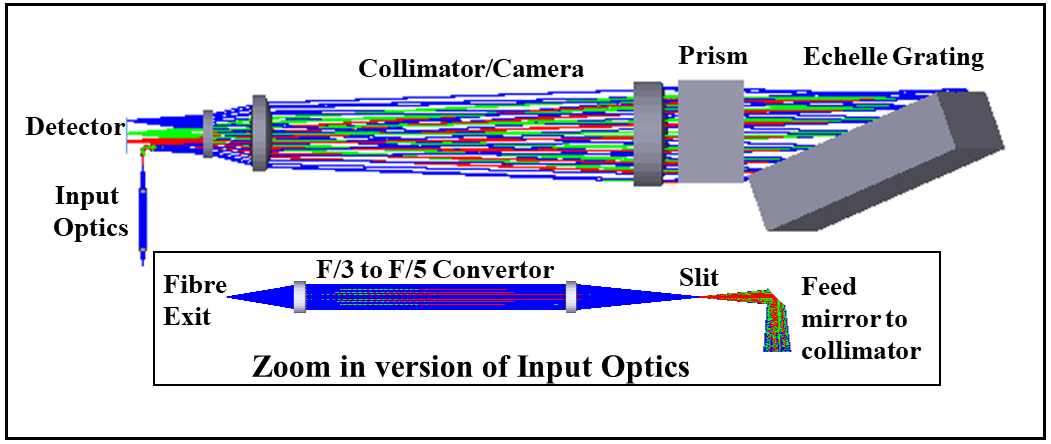}
	\end{tabular}
	\end{center}
   \caption
   { \label{fig:zmx_model} 
The optical design of the spectrograph developed in Zemax. A 100~$\mu \textrm{m}  $ (2.7 arc-sec) multi-mode optical fibre picks up the star-light from the prime focus of 2.3 $\textrm{m} $ primary mirror. The (f/3) beam emerging out from the fiber is converted to (f/5) beam in the spectrograph room, using Input Optics for f-ratio conversion. The f/5 beam is focused on the slit, the emerging beam from the slit is fed to the collimator with 45$^\circ$  feed mirror as shown in the zoom-in version of the optics. The collimated beam is forwarded to the prism (cross-disperser) and then to Echelle grating for horizontal dispersion. In the return path, the dispersed beam from the grating is passed through cross-disperser prism again. The dispersed beam is focused on the 4K$\times$4K CCD detector using collimator which also acts as a camera in the return path. }
   \end{figure}

The stellar light from the f/3 primary mirror of the telescope is picked up using 100~$\mu$m  multi-mode optical fiber. The exit of the fiber is fed to input optics (f/3 to  f/5 converter) that is used to focus light on the slit. The light from the slit is forwarded to f/5 collimator of focal length 755mm. The collimated light is directed to the cross-dispersing prism as shown in Figure \ref{fig:zmx_model}. The primary dispersion happens using Echelle grating in the horizontal direction. The reflective echelle grating redirects the high dispersed beam onto the cross-dispersing prism for the dispersion in the vertical direction. In the return path, the collimator acts as a camera and focuses the beam on to a 4K$\times$4K CCD detector. The Echelle grating is operated in Littrow mode with a blaze angle of ${\theta}${\tiny B}= $70\,^{\circ}$. 

\subsection{Grating position uncertainty}
\begin{figure}
   \begin{center}
   \begin{tabular}{c} 
   \includegraphics[scale = 0.4]{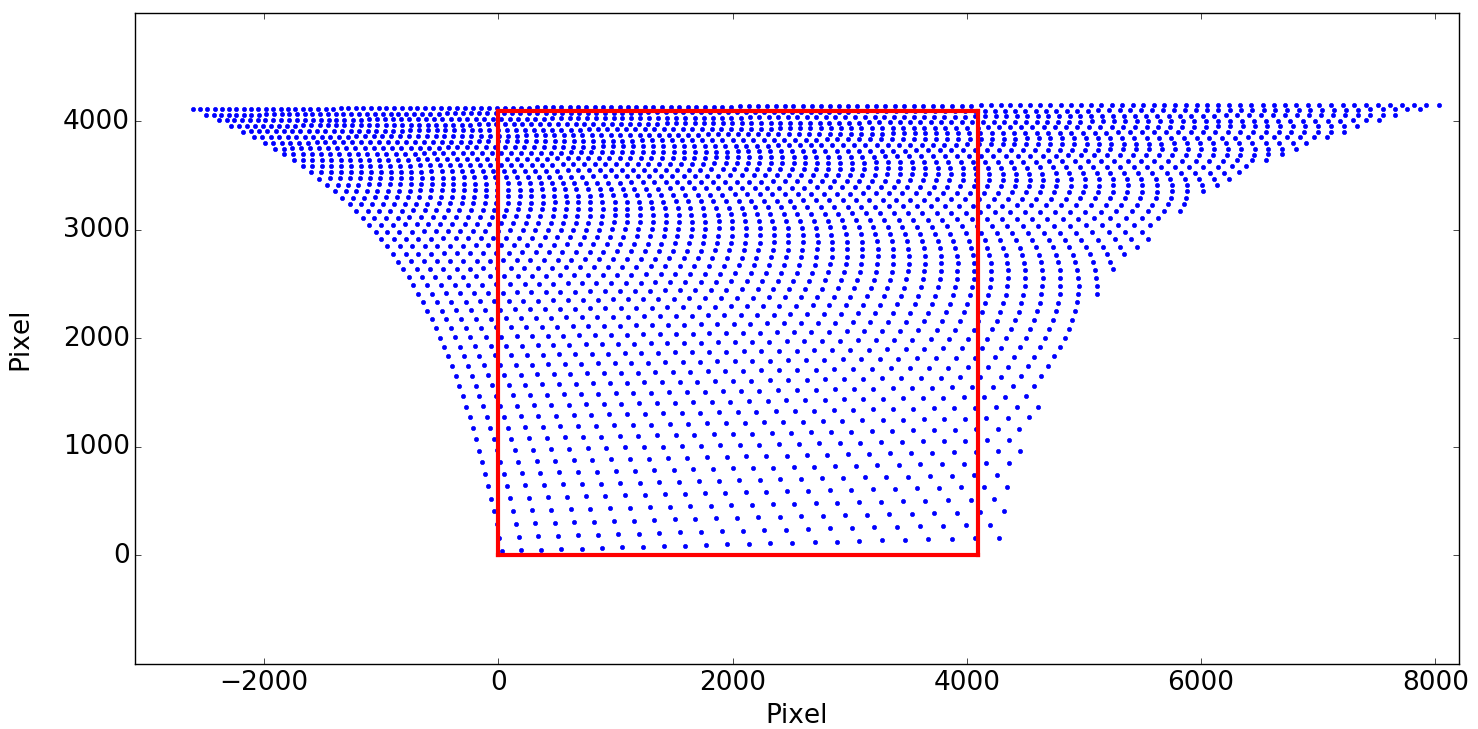}
	\end{tabular}
	\end{center}
   \caption
   { \label{fig:grating_gaps} 
   Echelle spectral format on the 4k~$\times$~4k CCD. Each point represents a single wavelength - range from 4000~\AA~  (bottom - 89$^{th}$ Echelle order) to 10,000~\AA~(top - 36$^{th}$ Echelle order). Each wavelength point is separated by 2~\AA~. In each order, the wavelength increases from left to right. The red box represents the spectral print on the CCD detector. For specific science cases, the wavelengths outside the detector format are recorded by shifting the entire spectra with grating rotation.
 }
   \end{figure}

The Echelle spectra (4000--10,000\AA) is recorded on a 4k$\times$4k CCD detector as shown in  Figure \ref{fig:grating_gaps} \cite{gratingmove}. At nominal setting, the spectral image overfills the detector plane.  Certain wavelengths on either side of each Echelle order,  fall outside the detector boundaries as shown in Figure \ref{fig:grating_gaps}. For specific science cases, the missing wavelengths at the edges are usually accessed by rotating the Echelle grating. The movement of the grating introduces uncertainties that invariably impacts the position repeatability, thus severely compromising the science that needs high RV precision. Since VBT Echelle is a general purpose spectrograph and used in a broad range of science observations, the grating needs to be moved to observe the wavelength of interest on a case-to-case basis. Due to contingent movement, it is difficult to maintain a well-defined zero-position of the grating. The problem is further aggravated by the backlash error in the grating-drive, non-uniformity of the gears and imperfect locking and holding. This would also result in inaccurate zero-positioning when the grating motor is turned off at the end of the observations on a day-to-day basis. With the instrumental restrictions, the grating has a positional uncertainty of $\pm  $ 25 arc-sec ($ \approx \pm $ 0.01 degree).     

\begin{figure}
   \begin{center}
   \begin{tabular}{c} 
   \includegraphics[scale = 0.4]{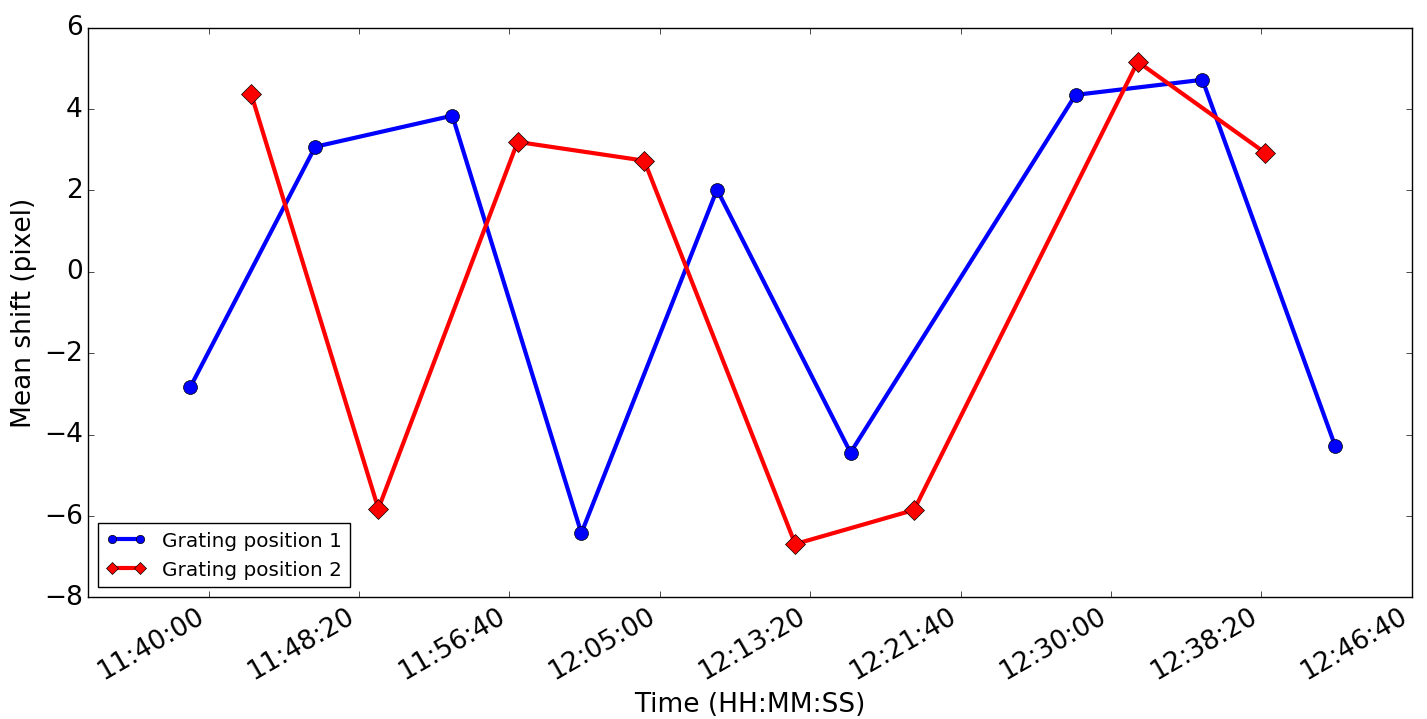}
	\end{tabular}
	\end{center}
   \caption
   { \label{fig:repeatability} 
The blue and red lines (each point represents the mean shift of all orders with respect to the reference position) show the pixel shifts obtained from Th-Ar frames taken repeatedly by moving the grating between two angular positions. The RMS scatter $\approx \pm $~4.5~pixels corresponds to position uncertainty of the movable grating.}
   \end{figure}

 We have used Th-Ar observations to estimate the repeatability of the calibration lines, which in turn affects the calibration consistency. The grating was commanded to move back and forth between two angular positions to determine the position repeatability. Between the moves, a Th-Ar spectrum was taken at each reference position and a total of 8 frames/position were obtained in about 1 hours duration. The first Th-Ar spectra for each position was cross-correlated with its subsequent frames to estimate the repeatability of the grating as shown in Figure \ref{fig:repeatability}. The plot shows a large RMS scatter of $ \pm $~4.5~pixels that can be attributed to the lack of position repeatability. It has also been analyzed that the shift is not the same from one order to another within the same spectrum. This intra order shift is estimated as 0.075~pixels. The lack of zero positioning of the grating is also noticed when the grating motor is turned off at the end of an observational run and turned on again for the next day observations. This would result in inconsistencies in the wavelength calibration procedure.

 \subsection{Vignetting}

\begin{figure}
   \begin{center}
   \begin{tabular}{c} 
   \includegraphics[scale = 0.45]{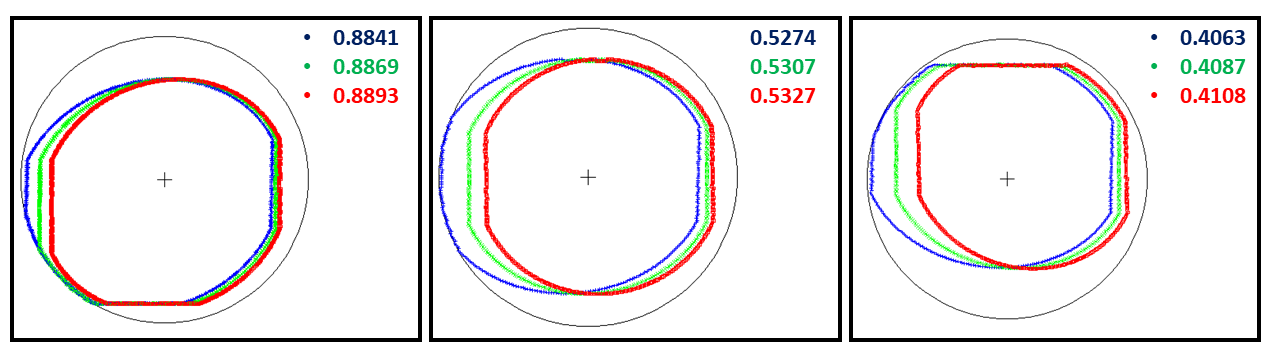}
	\end{tabular}
	\end{center}
   \caption
   { \label{fig:vignetting} 
Vignetting present in the spectrograph across the different Echelle orders as per the footprint diagram at the entrance of the camera. Three echelle orders - each order with two extreme and one central wavelength are evaluated.  The left plot shows the red region of the spectrograph (40$^{th}$ Echelle order), the vignetting from the prism effects the bottom part of the spectra. The middle plot shows the central region of the spectrograph (67$^{th}$ Echelle order), the vignetting effects only to the left and right side of the beam footprint (spectra) due to the grating. The right plot shows the blue region of the spectrograph (87$^{th}$ Echelle order), the vignetting effects the top part of the spectrograph. The corresponding wavelengths (units - micron) analyzed in each order is also displayed.}
   \end{figure}

 The spectrograph suffers from vignetting as the size of the grating and prism is smaller than the beam footprint (after the dispersion from grating) at different wavelengths.  We analyzed the vignetting effect in the system across the detector field using Zemax model of the spectrograph that is shown in Figure~\ref{fig:zmx_model}. Three Echelle orders, at the top, center, and bottom, are considered for evaluation. In each order, three wavelengths to the left, center and right are chosen. The wavelength selection is done in such a way as to sample the entire detector for vignetting effects. Figure~\ref{fig:vignetting} shows the vignetting present in our system in the form of truncated beam print at the entrance of the camera in the return path. In the absence of vignetting, the beam footprint is expected to be circular. However, due to the small size of the optics, there is vignetting at the edges of the image. Furthermore, the effect is found to be wavelength dependent across the orders and also varies from start to end of each Echelle order. Vignetting causes a change in the centroid of the spectral lines that fall on the detector. The effect is analyzed across the detector and is expected to give a constant outcome as it is a design constraint. However, as shown in Figure \ref{fig:repeatability}, due to uncertainty in zero-positioning, the landing of the spectra on the detector might vary. Thus the spectral lines that land at different locations on the detector experience a distinct vignetting effects, this results in the errors in centroid estimation of the spectral lines from spectrum to spectrum.  To estimate the vignetting from the design; the size of the grating, prism, and the collimator/camera optics have been increased to contain the entire footprint of the beam.  The centroid position with and without vignetting is compared and the error induced by vignetting was found to be $\approx$ 56 $\textrm{ms}^{-1} $. 
 
 \subsection{Distortion}
 A mathematical model of a 2D dispersion map of the spectrograph is constructed using grating and prism dispersion equations along with the spectrograph parameters (focal length of the collimator/camera, the location of the detector, etc) \cite{mathematical}. The isolated peaks of Th-Ar observational spectra are identified and the corresponding wavelength and the echelle orders are given as inputs to the mathematical model and Zemax model. The position of ThAr lines on the detector is estimated for both the models. Later, the models are perturbed in grating blaze angle by the same amount. The difference in which the models responds to the perturbation gives the resultant distortion in the optical design. Zemax model takes into consideration the inherent distortion present in the system. The difference in the positions of isolated ThAr line peaks, between the Zemax and the mathematical model, are estimated to account for the distortion. A residual from the comparison is a measure of distortion which was found to be 400~$\textrm{ms}^{-1}$.

 \subsection{Other factors}
Apart from vignetting and distortion, there is also tilt in the spectral output at the detector. The aberrations in the spectrograph would result in the asymmetry of the Point Spread Function (PSF), degrading the resolution at the detector \cite{PSFvbt}.  Also, the spectrograph is not vacuum stabilized, hence the environmental variations are expected to affect the observations. The inherent stability of the spectrograph over several days observations (without  the grating movement) was estimated to be about $ \pm $ 1 pixel ($ \pm $ 1000 $\textrm{ms}^{-1} $) \cite{stabilityVBT}. 

\section{Necessity of correcting for the zero position drifts}
\label{sect:need}

 ThAr lamp is the wavelength calibration source in the spectrograph. The calibration spectra are currently taken either before or after the stellar observations. The wavelength calibration is done as per the conventional procedure of generating a template spectrum with the spectrograph. The ThAr lines are identified and wavelength calibrated for a well-exposed spectrum taken with the spectrograph. This spectrum is used as a template for wavelength calibration. Rest of the ThAr frames obtained as part of a regular observational run are correlated with the template spectrum to obtain the dispersion solution. In the standard calibration procedure, the ThAr spectrum is correlated with the template to determine the overall shift w.r.t each order (or entire spectrum). If there is a shift in the observed spectrum from correlation result, then the spectrum is corrected for the shift and the dispersion solution is applied. The shift applied is often a constant number which does not take into account the non-linearities along and across the spectral orders. This procedure might result in a discrepancy in the wavelength solution in the following ways \cite{model_based, HESP}:
 \begin{itemize}

\item The transient distortions to the local wavelength solutions can be corrected/tracked by wavelength calibration sources if the shifts in the calibration line positions are relatively small (at ±1 pixel level or less). In our case, the grating repeatability is uncertain up to $\pm$4.5 pixel level as shown in Figure \ref{fig:repeatability}. Such large perturbations need to be properly accounted for and corrected.

\item The blends in the ThAr calibration lines are usually identified by fitting a single Gaussian to measure the centroid. However, due to a shift of ± 4.5 pixel, these lines experience different aberration effects as they pass through the optical components. This leads to inaccuracies in the centroid estimation.  

\item As explained in Section \ref{sect:limitations}, apart from the uncertainty in zero-positioning, there are other effects like vignetting, distortion and environmental instabilities in the spectrograph. This might lead to non-linear effects on the spectral shift across the dispersion and cross-dispersion direction. We also find that wavelength shift across the orders, due to large perturbations, do not remain uniform. The conventional correlation of ThAr spectrum with the template would not take into account these non-linearities.

\end{itemize}
 
 The echelle spectrograph operating at VBT lacks the mechanism to revert the optical components to their original position after the movement. In such cases, it is essential to correct for the positional drifts. After the perturbations in the optics are corrected, the obtained spectra can then be correlated with the template for wavelength calibration. 
 
 We propose to use a ThAr spectrum taken at the beginning of the observational run to estimate the perturbations in the positions of the components. The centroid positions of the isolated peaks in ThAr spectra are estimated. These centroid locations depict the current positions of the optical components in the spectrograph. If there are any perturbations in the optical components, then there will be shifts in the centroid positions of the isolated wavelengths. 
 
 \subsection{Using Zemax model to  estimate instrumental drifts}
\begin{figure}
   \begin{center}
   \begin{tabular}{c} 
   \includegraphics[scale = 0.6]{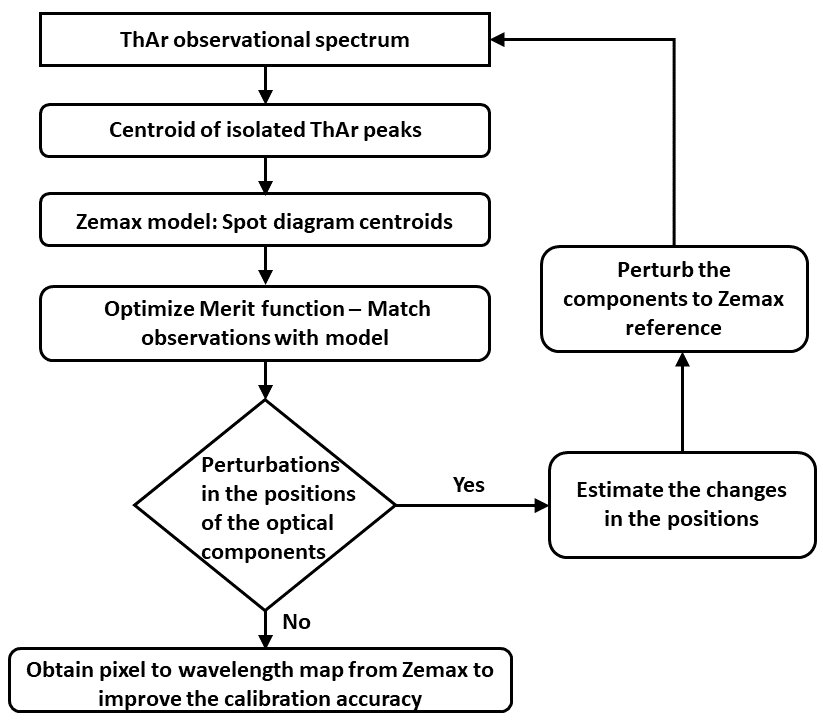}
	\end{tabular}
	\end{center}
   \caption
   { \label{fig:concept} 
Flow chart of model matching with the observations.}
   \end{figure}

  To estimate the optomechanical positional variations of the components, we have used a Zemax based ray trace model. The flow chart of the approach is demonstrated in Figure \ref{fig:concept}.  Wavelengths of isolated Th-Ar peaks from observational spectra are identified; the wavelength and the corresponding Echelle order is fed to Zemax using Python interface. We have used the Spot diagram (CENX and CENY) of isolated ThAr wavelength from Zemax for evaluation of the centroid shift as described in (Grupp et al, 2009)  \cite{tandp}.  The Zemax parameters are perturbed in such a way that the centroid positions (Spot diagram) of the Zemax model are matched with the Th-Ar line centers obtained from the ThAr observations \cite{model_based}. In contrary to the traditional model based calibration approaches, we have used a Zemax optical model of the spectrograph instead of the physical model of the instrument. For this purpose, a merit function is constructed to allow perturbations of RV sensitive parameters, such that the Zemax model would match with the observations after optimization of the merit function. After the model is matched with the observations, the perturbations in the positions of the components are estimated.

Since the error-free locking is not ensured for the optical components, we are using reference locking for precision RV observations. Zemax model is used as the reference and all the optical components are brought to the reference level at the beginning of each observational run during the day time. We also propose that the Zemax model can be used to improve the calibration accuracy after the match is obtained with the observations.   The primary calibration is always done with the Th-Ar source in the spectrograph. However, Th-Ar lines are not distributed uniformly and it also has some saturated and blended lines. This has an impact on the accuracy of the wavelength solution. The pixel-to-wavelength map obtained from the Zemax model could improve the wavelength solution. The dispersion solution can be extended to the region where ThAr lines are inadequate \cite{HESP_performance}. A wavelength calibration process assigns a unique wavelength to each pixel on the CCD. It is a crucial data reduction step for all spectroscopic measurements. The error induced in the wavelength calibration can significantly affect the precision obtained in the RV observations. The model-based approach can be extended to obtain the wavelength-pixel map for features across the spectrographs using interpolation, thus improving the RV accuracy of the spectrograph.

 The approach we propose to improve the calibration accuracy is as follows: The first step is to estimate the wavelength – pixel positions of isolated ThAr lines from observations. These wavelengths along with the echelle order are to be given as input to Zemax. The Zemax model determines the position of the wavelength on the CCD as per the instrumental design along with aberrations and perturbations. From the wavelength-pixel map obtained from both ThAr observations and Zemax model, in each order, one wavelength position (which is the best match between the model and observation) is taken. This location is anchored and the relative position (between Zemax and observations) of the other wavelengths in each order is obtained. Then using the relative positions, the wavelength-pixel map is interpolated for locations where ThAr lines do not exist.  
 
Apart from that, traditional calibration techniques have the limitation of stability and lifetime issues. The RV observations of exo-planets, e.g.,  typically has a  timeline ranging from weeks to years. In such a case, it is essential to ensure that the wavelength calibration errors are minimized. The model-based approach can be used to establish the accuracy using the real-time setup of the spectrograph and the Zemax model. Wavelength calibration is also unique to each spectrograph and the model-based approach  can be extended to improve the calibration procedure and the RV accuracy.

\subsection{Degeneracy in the model}
The effect of degeneracy in the model for estimating the positional variations is also evaluated to verify the credibility of the model matching. As part of this analysis, optical components are perturbed with a known value. The system is optimized to bring the components back to the nominal position. We then estimated the difference in predicted vs actual (perturbed) positions. Table \ref{tab:degeneracy} shows that the degeneracy present in the system is minimal compared to the accuracy we are targeting in model matching (0.01 degree). The uncertainties in positioning the components in the spectrograph are to a range of 0.01 degree.  The spectrum that is used as the input to the model is obtained from the spectrograph and is not a synthetic spectrum. Thus some inherent noise is expected in the spectrum even after the initial data reduction was carried out using IRAF. However, to evaluate the uniqueness of the model in predicting the perturbations of the optical components, we have added noise in the spectrum. The radial velocity error obtained in the comparison of the noise added spectrum to the original spectrum is 7 $\textrm{ms}^{-1}$. The echelle spectrograph operating in VBT is a general purpose instrument and is not designed for precision radial velocity measurements. Thus, the stability limitation is around 50 $\textrm{ms}^{-1}$ with the spectrograph. The model deviation is within the stability limits of the spectrograph.

\begin{table}
\begin{center}
\begin{tabular}{|l|l|l|l|}
\hline
\textbf{Perturbed parameter} & \textbf{Actual parameter values} & \textbf{Values after optimization} & \textbf{Difference} \\ \hline
prism\_x                     & 8.585863342                      & 8.585863042                        & -3E-07              \\
prism\_y                     & 0                                & -5.39281E-06                       & -5.39281E-06        \\
prism\_z                     & 0                                & 1.06121E-06                        & 1.06121E-06         \\
grating\_x                   & 25.852965                        & 25.85295965                        & -5.35E-06           \\
grating\_y                   & -68.555                          & -68.55499933                       & 6.7E-07             \\
grating\_z                   & 90                               & 89.99999423                        & -5.77E-06           \\ 
detector                     & 0                                & 4.65131E-07                        & 4.65131E-07         \\ \hline
\end{tabular}
\end{center}
 \caption
   { \label{tab:degeneracy} 
The perturbed parameter values compared with the reference positions. The "difference" column shows the difference between the reference and the perturbed positions after optimization of the merit function. In the view of estimating the degeneracy in the system, the decimal positions are increased. }
\end{table}   

\section{Sensitivity Analysis in Zemax}
\label{sect:sensitivity}  

\subsection{Procedure}
The prerequisite of modeling the observations in Zemax is to estimate the RV sensitivities of the optical components \cite{tolerancing}. For matching observations with Zemax model and construction of \emph{merit function}, the sensitivity of the optical components in echelle dispersion direction is estimated \cite{tolerancing}. 

\begin{figure}
   \begin{center}
   \begin{tabular}{c} 
   \includegraphics[scale = 0.4]{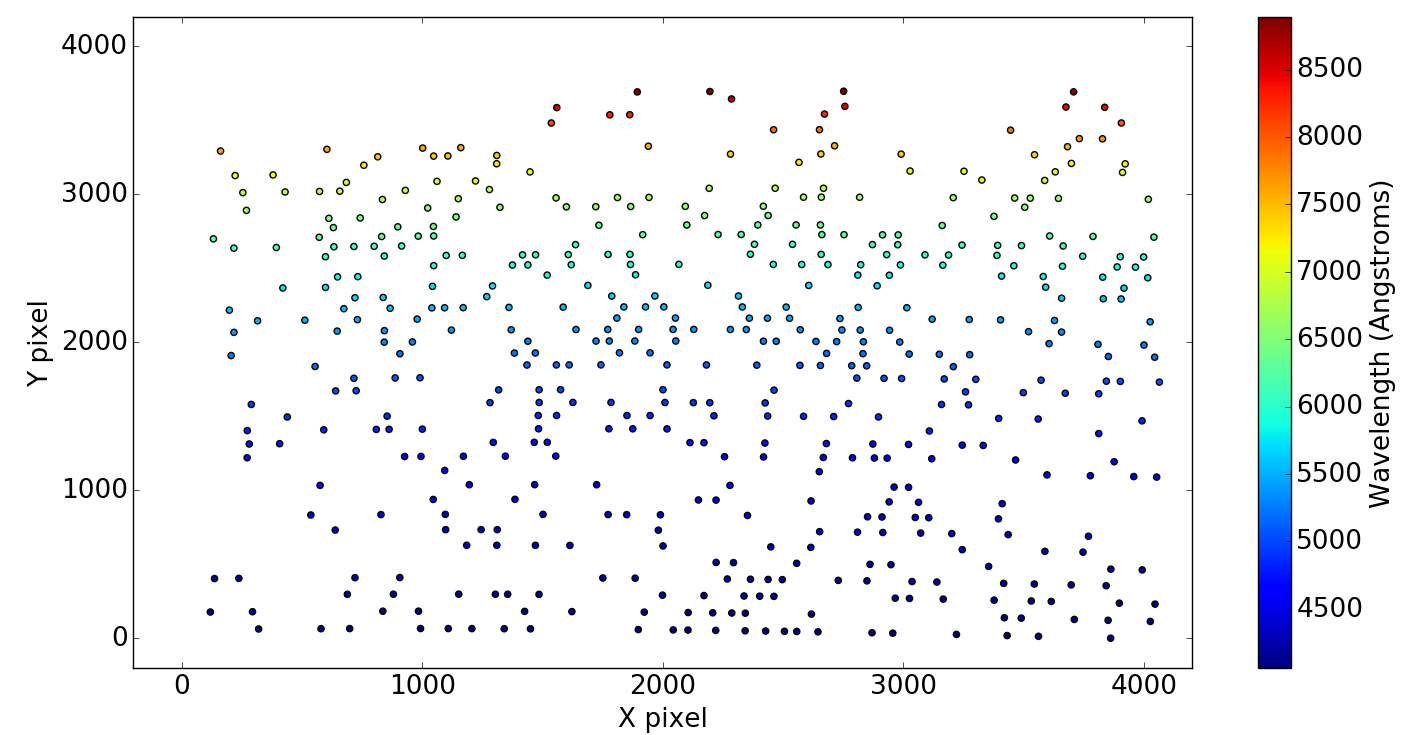}
	\end{tabular}
	\end{center}
   \caption
   { \label{fig:wavelength} 
Wavelength map of the isolated Th-Ar lines (marked as dots)  across the 4k$\times$4k CCD. There are 50 Echelle orders in the spectrograph with a wavelength range from 4000 -- 10000 \AA. In the sensitivity analysis, a total of 500 isolated wavelengths were used. The color bar to the right shows the wavelength mapping from blue to red.
}
   \end{figure}

A Th-Ar spectrum is taken for sensitivity analysis and the wavelength position of the isolated peaks is determined. Figure~\ref{fig:wavelength} shows the isolated wavelengths identified from the Th-Ar reference frames. These wavelengths and the corresponding Echelle orders are given as input to Zemax using a Python interface. Coordinate breaks in Zemax are placed in such a way that each optic is perturbed separately about the local axis of the component. Then a known perturbation (in positive and negative direction separately) is applied to each optical component in the Zemax model and the perturbed X, Y centroid position is estimated  \cite{zemaxmatlab}.

A common practice is to consider the translational and rotational degrees of freedom for sensitivity analysis. However, the optical components (prism, grating, etc.) of the spectrograph are in the collimated beam, and they are less sensitive to translational perturbation in $X, Y, Z$ directions. Hence, we have analyzed only rotational perturbations. A tilt of $\pm$0.01$^\circ$ in $X, Y,$ and $Z$ direction is considered for this analysis. The rotational perturbations with respect to the slit are given as follows:
\begin{itemize}
\item $X$ perturbation: Movement across the slit length
\item $Y$ perturbation: Movement across the slit width
\item $Z$ perturbation: Movement across the normal to the slit.
\end{itemize}

\subsection{Results - RV sensitivity from individual components}

The isolated wavelengths obtained from Th-Ar observational frame are incorporated into sensitivity analysis. The nominal $X$ centroid position of the optical component is compared with the perturbed position from the spot diagram. The results of the sensitivity analysis for the perturbation of grating, prism, input optics, collimator, and feed-mirror are summarized in  Table \ref{tab:perturbations}. In the table, the maximum to the minimum pixel shift from the nominal positions for individual perturbation is listed.

\begin{figure}
   \begin{center}
   \includegraphics[scale = 0.45]{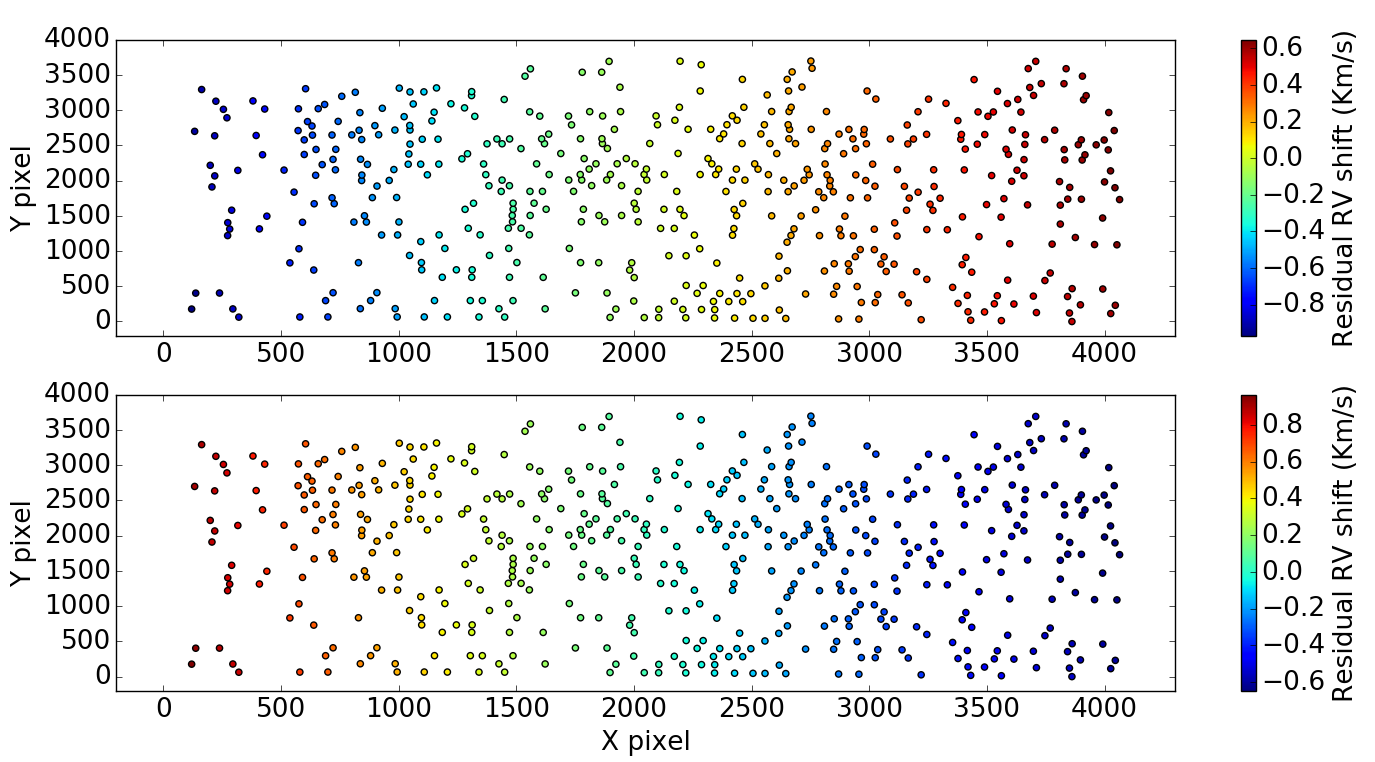}
	\end{center}
   \caption
   { \label{fig:grating_Yrv} 
The difference in $X$-centroid position for the perturbed (in the blaze direction) and the unperturbed system. The RV changes are relevant only in dispersion direction, therefore, spectral movement in $Y$-direction is not considered. The top panel shows positive perturbation of 0.01~$^\circ$ from the nominal position.  The $X$-centroid shifts are from the left to right across the Echelle order. If the grating position is uncertain to 0.01~$^\circ$, then the resultant residual shift is 1.4 pixels (1.4 $\textrm{kms}^{-1}$). The bottom panel shows the negative perturbation (-0.01$^\circ$) for the angle of incidence on the grating.  The comparison of both the plots shows that positive and negative perturbation causes equal but an opposite shift of lines across the detector. }
   \end{figure}

The Echelle grating shows the maximum sensitivity in the dispersion direction. A $\pm$0.01 $^\circ$ perturbation of grating angle of incidence from the nominal position produces a shift up to $\approx 20 $ pixels across different echelle orders. In the current optomechanical design of the spectrograph, the grating position uncertainty is about 25 arc-sec ($\approx$ 0.01$^\circ$), thus making this analysis essential. The measured RMS scatter of $\pm 4.5$ pixels from the grating uncertainty shown in Figure \ref{fig:repeatability} is well within the range predicted from the sensitivity analysis. From the spectrograph dispersion, 1 pixel shift in velocity units corresponds to $\sim1 \textrm{kms}^{-1}$.  Figure~\ref{fig:grating_Yrv} shows the motion of the spectra on the detector due to grating perturbation along the dispersion direction. The residual shift of the spectra after removing the mean motion (-20.95 pixels) corresponds to RV sensitivity of 1.4 pixels (1.4 $\textrm{kms}^{-1}$). In other words, angular uncertainty of 0.01 $^\circ$ in placing of the grating would result in a residual shift of 1.4 pixels. These effects should be carefully accounted for to improve RV precision. The matching of observations with Zemax model and estimating the positions of the optical components and bringing them to a known reference is intended to accomplish this goal.

\begin{table}
\begin{center}
\begin{tabular}{|l|l|l|}
\hline
\textbf{Component} & \textbf{Perturbation} & \textbf{Pixel shift ($ \textrm{kms}^{-1} $)} \\ \hline
Collimator         & Tilt $X$ (0.01)         & 0.6 to -0.8                      \\
Collimator         & Tilt $Y$ (0.01)         & -16.95 to -18                    \\
Feed mirror        & Tilt $X$ (0.01)         & 0.15 to -0.20                    \\
Feed mirror        & Tilt $Y$ (0.01)         & -0.735 to -0.855                 \\
Grating            & Tilt $X$ (0.01)         & 1.0 to -0.6                      \\
Grating            & Tilt $Y$ (0.01)         & 21.8 to 20.4                     \\
Grating            & Tilt $Z$ (0.01)         & 0.6 to -1                        \\
Input Optics       & Tilt $X$ (0.01)         & 0.045 to -0.045                  \\
Input Optics       & Tilt $Y$ (0.01)         & 0.56 to 0                        \\
Prism              & Tilt $X$ (0.01)         & 0.06 to -0.06                    \\
Prism              & Tilt $Y$ (0.01)         & 0.78 to 0.62                     \\
Prism              & Tilt $Z$ (0.01)         & -9 to -10.05                     \\
Temperature        & 20+0.5 deg         & 0.12 to 0                        \\
Pressure           & 1 + 0.12 Pa           & 0.48 to -0.08                    \\ \hline
\end{tabular}
\end{center}

 \caption
   { \label{tab:perturbations} 
Perturbations applied to the individual component for sensitivity analysis of the spectrograph. After the individual component is perturbed, the spot diagram of the perturbed system is compared with the nominal position. The maximum to the minimum range of the pixel shift is obtained across the detector. This range is displayed in the last column of the table. }
\end{table}

\section{Modelling Th-Ar observations in Zemax}
\label{sect:model}
\subsection{Procedure for model matching with observations}

The sensitivity analysis was done for individual components with a perturbation in one of the rotational degrees of freedom. However, in the actual working system, the perturbations could arise from other degrees of freedoms. This cumulative effect would result in shifting the ThAr line positions from the nominal location. We use ThAr observations taken at the beginning of the observational run on each day to estimate the perturbations in the optical components. The initial data reduction and wavelength calibration of Th-Ar data are done in IRAF. The latter part of the analysis is done using a dedicated pipeline developed in Python.

In this context, a \emph {merit function} was constructed using RV sensitive parameters (that can be perturbed in real time), to achieve the target position of the wavelength from Th-Ar observations.  The merit function constructed in such a way that the positions of the optical components (to be perturbed) are declared as variables; the target is the observational $X$-centroid position and the value of the merit function is the spot diagram CENX. Coordinate breaks are applied to perturb the positions of the optical components during optimization in the Zemax model.

In the analysis, we have considered the non-uniform illumination of the pupil plane\cite{apod1,apod2}. The telescope beam is picked up by a circular fiber and the starlight is fed to the spectrograph. To simulate this non-uniform illumination, we have used a Gaussian apodization factor 1 in Zemax (fall in the intensity of 13\% of the peak value). Apodization factor 1 was considered based on the analysis of beam exit from the fiber (exit pupil). The optimization of the \emph {merit function} is obtained by perturbing the three rotational degrees of freedom $(X, Y, Z)$ of the selected optical components.

\begin{figure}
   \begin{center}
   \begin{tabular}{c} 
   \includegraphics[scale = 0.4]{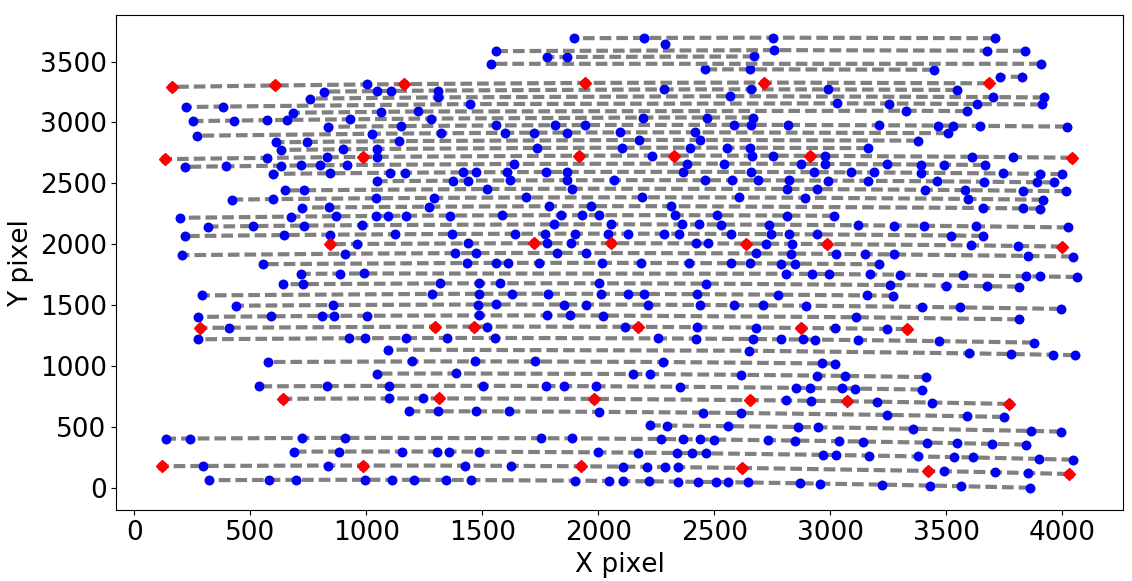}
	\end{tabular}
	\end{center}
   \caption
   { \label{fig:select} 
Comparison of the Th-Ar wavelengths used in the sensitivity analysis (blue-dots) and the wavelengths selected (red-diamond) for modeling the observations. The 50 Echelle orders are shown by dashed gray lines.  Six Echelle orders are considered and from each order six wavelengths (red-diamond) are chosen for the analysis. }
   \end{figure}

To reduce the computational load/intensity, in the initial case of a matching model with the observations, the wavelength selection is done such that sampling across the detector is maintained. Six Echelle orders are selected across the detector with well-sampled six wavelengths from each order across the FSR of each order.  Figure~\ref{fig:select} shows the wavelengths marked in red dots for the modeling the observations using merit function. They are compared with the wavelengths (marked in blue dots) used in sensitivity analysis as discussed in Section \ref{sect:sensitivity}. This sampling of $6 \times 6$ matrix of well-sampled wavelengths is used for estimating the perturbations. To evaluate the match of the model with the observations all the isolated wavelengths used in the sensitivity analysis are utilized.

\subsection{Results of model matching with observations}
\begin{figure}
   \begin{center}
   \begin{tabular}{c} 
   \includegraphics[scale = 0.4]{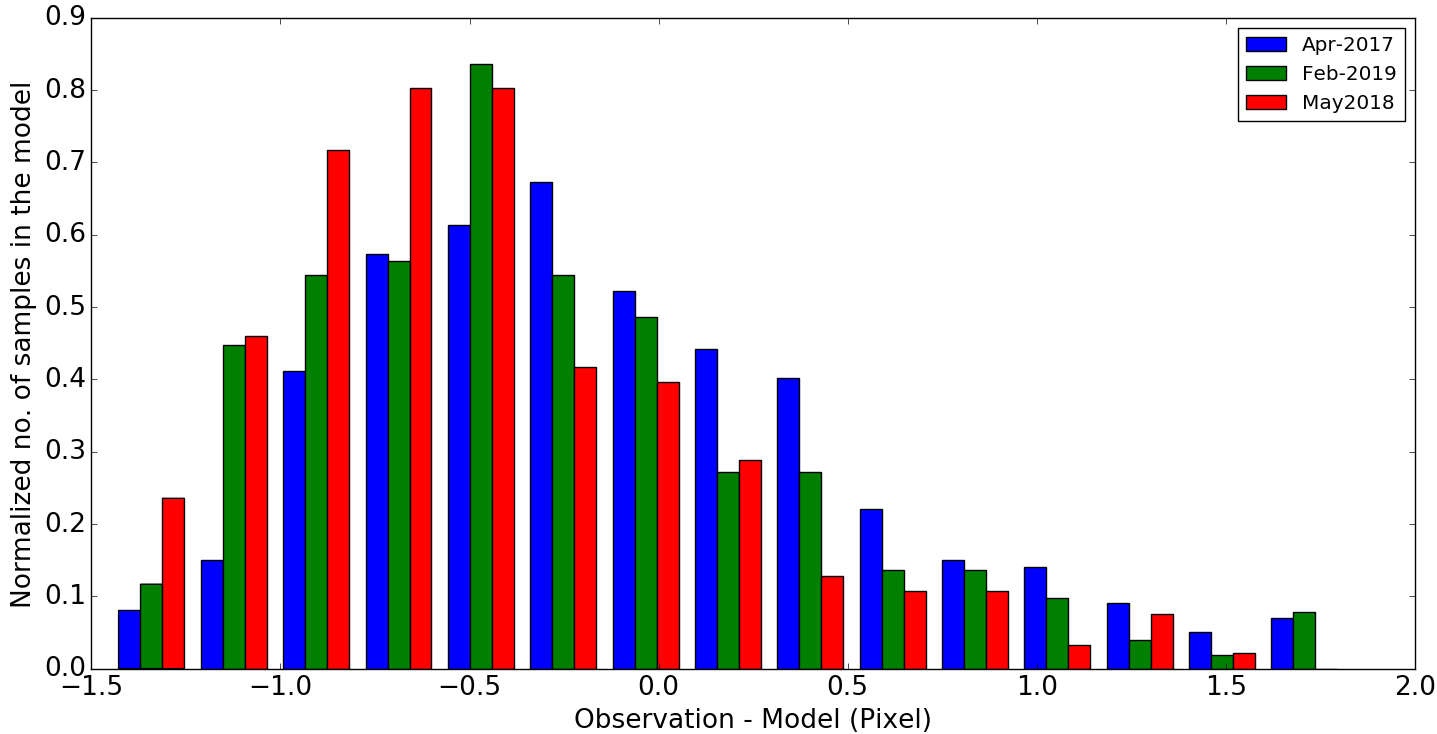}
	\end{tabular}
	\end{center}
   \caption
   { \label{fig:Model_match} 
   Histogram of difference between the model and  Th-Ar observations for the X-centroid position on the detector. Three ThAr observations taken at different time are used for matching the model. The plot shows that the model can match the observations with $\pm $ 1-pixel accuracy. }
   \end{figure}

The perturbations obtained from the match of a model with the observations are given as input to a native Zemax model. For the selected 500 Th-Ar lines (as discussed in  Section \ref{sect:sensitivity}), the X-centroid is computed in the perturbed Zemax model. These centroid positions are compared with the observational positions. The best match is within $\pm$ 1-pixel accuracy as shown in Figure \ref{fig:Model_match}. The model is validated by obtaining the match to a different set of ThAr frames. The positional parameters of the components might vary from the actual system parameters during the built and alignment phase. The parameters as per the design phase have a deviation of $\pm$ 24 pixels from the observations. After modeling and optimization, the match between the observation and the model is obtained as $\pm$ 1 pixels.

The goodness of the model also depends on the wavelength identification of Th-Ar observational spectra, as the wavelength and the corresponding echelle orders are used as inputs for the Zemax model. Th-Ar spectrum has few saturated lines \cite{ThAr} and the wavelength identification has an inaccuracy in the case of saturated lines. In the model matching, we have eliminated the outliers of saturated lines beyond the wavelength range of 7000~\AA. 
 
\section{Conclusion}
\label{sect:conclusion}  

With these current settings of the spectrograph, to achieve a radial velocity precision of 10 $\textrm{ms}^{-1}$, the instrumental stability has to be at 0.01-pixel level.  For instance, the grating position has ambiguity $\approx$ 25 arc sec, during the system restart for everyday observations.  This results in a residual shift of 1.4 pixels across the Echelle orders. In the current configuration, the grating uncertainty could very well impede the achievable precision of the instrument. Apart from this, the design of the spectrograph suffers from vignetting (56 $ \textrm{ms}^{-1} $) and distortion effects (400 $ \textrm{ms}^{-1} $).  The vignetting and distortion that could affect the RV precision were not taken into account during the design phase in the instrument as it was not designed as an RV spectrograph. The temperature and pressure instabilities in the spectrograph room results in RV shift of 120 $ \textrm{ms}^{-1} $ and 400 $ \textrm{ms}^{-1} $, respectively. In real time, due to the cumulative effect of these factors, the overall achievable RV precision is compromised. Hence it is essential to correct the day-to-day drifts in the spectrograph using Zemax model as proposed.

 Apart from estimating instrumental perturbations, Zemax model can be applied to improve the calibration accuracy. In the view of achieving improved calibration accuracy,  ThAr observations are to be taken before and after the star of interest observations \cite{bracketting}. If there are any drifts due to instrumental perturbations, these drifts would be on a linear scale (instrumental drifts are estimated as a linear trend in short period of 10-15 minutes). The instrumental shift is assumed to be on a linear scale in comparison to the pixel/wavelength shift which has non-linearities. Both the ThAr exposures are evaluated for the perturbations in the optical positions and we use the model to estimate the wavelength solution of the target star observation in between the ThAr exposures. This would improve wavelength calibration accuracy and can also be used to estimate the stability of the spectrograph using the ThAr exposures before and after star observation.  However, if the Zemax model is matched with the observations, the model takes into account the non-linearity and the aberrations in the system to improve the calibration accuracy.

 This technique can also be applied for highly stable Radial velocity spectrographs targeting for few $ \textrm{cms}^{-1} $ precision measurements. Using the model-based approach, the accuracy of the wavelength calibration in case of simultaneous calibration can also be improved \cite{HESP_performance}. However, this technique does not require the physical modeling of the spectrographs but can employ the commercially available Zemax platform for model matching. As the Zemax model can be used to predict the instrumental perturbations, the shifts in the instruments can be restricted. This can also be applied in the case of a frequency comb or any known calibration lines to improve the accuracy of the model based calibrations.

In conclusion, we have analyzed a general purpose spectrograph with movable components and proposed a methodology to predict and correct the position of the components at the beginning of the RV observations.  Understanding the attainable RV precision would give an estimate on required up-gradation of the optomechanical mounts and the environmental control. 

\section{Acknowledgements}

Authors would like to thank the technical and observing staff at Vainu Bappu Telescope. We also acknowledge the valuable support and suggestions received from Dr. Sivarani Thirupathi during this work.  Finally, we would also like to thank referees for their critical review and useful feedback that helped us to improve the manuscript.


\bibliography{report}   
\bibliographystyle{spiejour}   

\listoffigures
\listoftables

\end{spacing}
\end{document}